\documentclass[pra,preprint,superscriptaddress,showpacs]{revtex4}
\usepackage{times}
\usepackage{amsbsy,amssymb}
\usepackage{graphicx,color}

\newcommand{\ket}[1]{|{#1}\rangle}
\newcommand{\bra}[1]{\langle{#1}|}

\newcommand{\id}{I}

\newcommand{\Comment}[1]{}
\newcommand{\Oldver}[1]{}

\begin{document}

\title{Liquid state NMR simulations of quantum many-body problems}

\author{C. Negrevergne}
\thanks{Corresponding author. Email: camille@iqc.ca}
\affiliation{Institute for Quantum Computing, University of Waterloo, Waterloo,
ON, CANADA N2L 3G1}

\author{R. Somma}
\affiliation{Los Alamos National Laboratory, Los Alamos, NM 87545, USA }
\affiliation{Centro At\'omico Bariloche and Instituto Balseiro,8400 San Carlos 
de Bariloche, Argentina}

\author{G. Ortiz}
\affiliation{Los Alamos National Laboratory, Los Alamos, NM 87545, USA }

\author{E. Knill}
\affiliation{ Math. and Comp. Sciences Div., National Institute of Standards 
and Technology, Boulder, CO 80305, USA}

\author{R. Laflamme}
\affiliation{Institute for Quantum Computing, University of Waterloo, Waterloo,
ON, CANADA N2L 3G1}

\date{\today}

\begin{abstract}
  
  Recently developed quantum algorithms suggest that in principle,
  quantum computers can solve problems such as simulation of physical
  systems more efficiently than classical computers. Much remains to be
  done to implement these conceptual ideas into actual quantum
  computers. As a small-scale demonstration of their capability, we
  simulate a simple many-fermion problem, the Fano-Anderson model,
  using liquid state Nuclear Magnetic Resonance (NMR). We carefully designed
  our experiment so that the resource requirement would scale up
  polynomially with the size of the quantum system to be simulated. The
  experimental results allow us to assess the limits of the degree of
  quantum control attained in these kinds of experiments. The
  simulation of other physical systems, with different particle
  statistics, is also discussed.

\end{abstract}

\pacs{03.67.-a, 05.30.-d, 76.60.-k, 03.65.Yz} 

\maketitle

\section{Introduction}
\label{introduction}

Quantum mechanical systems provide new resources to solve problems
which are difficult to solve on classical computers. If we had a large
quantum computer today, we could break cryptographic codes \cite{shor},
perform a variety of search algorithms \cite{grover,ambainis}, estimate
eigenvalues of operators \cite{kitaev,ekert}, or simulate quantum
systems \cite{feynman}. In particular, the latter would enable a
better understanding of the quantum world
by enabling analyses of complex chemical
reactions or demonstrating new states of matter.
However, questions like What are the physical quantum states that can be
reached efficiently? or  What kind of physical processes can be
efficiently simulated on a quantum computer? still remain open.

Since Richard P. Feynman conjectured that an arbitrary discrete quantum
system may be {\it simulated} by any other \cite{feynman}, the
simulation of quantum phenomena became a fundamental problem that a
quantum computer, i.e., a universally controlled quantum system, may
potentially solve in a more efficient way than a classical computer.
The basic idea is to imitate the evolution of a physical system by
cleverly controlling the evolution of the quantum computer. Quantum
simulation is the process of faithfully imitating a physical phenomenon
using a quantum computer. Although Feynman's illuminating conjecture
seems appealing, it was only recently proved generally valid
\cite{note1, ortiz1, batista1, somma1, batista2}. Experimentally
demostrating that one has universal control and thus can quantum imitate an
arbitrary physical process constitutes an extremely challenging
enterprise. 

It is important to notice that the efficiencies of quantum simulating
the evolution of a physical system and of obtaining the sought-after
information about a physical property must be established separately
in most cases. A demonstration that evolution can be simulated
efficiently~\cite{slloyd, ortiz1, somma1, batista2}, that is, can be
simulated with polynomial resources as a function of problem size, is
in general insufficient for showing that the desired property (e.g.,
the ground state energy of a given Hamiltonian) can be obtained
efficiently also.  In general, the exponentially large Hilbert space
that characterizes those physical systems and the inherent quantum
parallelism of a quantum computer are insufficient for showing 
that an algorithm for
quantum computation efficiently solves a problem.  We pointed out
in~\cite{ortiz1, somma1} that in a quantum computation, it is
necessary to demonstrate that in addition to maintaining adequate
accuracy (noise, approximations, and statistical error control) one
also has to demonstrate the polynomial scaling of the three main steps
of a simulation, initialization, propagation and measurement.

Some quantum processes can be simulated very well and efficiently on
classical computers. Simulating quantum phenomena using stochastic
approaches reduces the problem to quadratures, which are
multidimensional integrals that can be computed using Monte Carlo
techniques.  In general, the complexity of deterministic
$N$-dimensional integration is of order $\varepsilon^{-N/\alpha}$
(i.e., exponential in $N$), where $\varepsilon < 1$ is some stipulated
error and $\alpha$ quantifies the smoothness of the integrand. On the
other hand, the expected complexity of Monte Carlo integration is of
order $\varepsilon^{-2}$, and hence independent of $N$ and
$\alpha$ (assuming that the variance of the integrand is finite).
The reason for introducing these statistical techniques was
to overcome the exponential complexity of deterministic approaches
such as the Lanczos method \cite{lanczos}.  Realistic models of liquid
or solid $^4$He have been simulated to experimentally measured
precision for a few years \cite{ceperley}. Recently developed
loop-cluster algorithms allow highly efficient and informative
simulation of many quantum spin models of magnetism \cite{cluster}.

An important class of problems for which classical computers have major
difficulties is the simulation of interacting fermionic systems (almost
all {\it large-scale} simulations of fermions are done by the Monte Carlo
method). In fact, as noted in~\cite{ortiz1,somma1}, Feynman and others
prior to him intuited this difficulty. Unless an approximation is made,
the various quantum Monte Carlo algorithms must inevitably sample from
a multivariate distribution $P$ that has regions of phase space where
it is negative that are comparable to regions where it is positive
(because the state function belongs to the totally antisymmetric
representation of the permutation group). In general, the nodal
hyper-surface $P=0$ separating the regions is unknown (an exception
being when symmetry considerations alone determine it), making it
impossible to solve the problem by independently sampling from each
region where $P$ has a definite sign. The sign problem is prohibitive
on a classical computer because it results in the variance of
measured quantities growing exponentially with the number of degrees of
freedom of the system. Still other applications require sampling from a
complex-valued distribution $P$. This occurs, for example, if the
simulation is done as a function of real Minkowski time or if
time-reversal symmetry is broken. In previous
work~\cite{ortiz1,somma1}, we have discussed how certain sign problems
can be overcome using quantum network algorithms.

In this paper we describe how quantum simulation of many-body problems
can be realized in liquid state NMR Quantum Information Processors
(QIPs) \cite{NMR1}. The constituents of the system may represent
particles with arbitrary exchange statistics and generalized Pauli
exclusion principle (such as fermions obeying Fermi statistics), spins,
etc. In particular, we show how to efficiently imitate a resonant
impurity (localized state) scattering process in a metal (which is made
of fermions), using the nuclear spins of a trans-crotonic acid
molecule.  This problem is physically modeled by a Fano-Anderson
Hamiltonian \cite{ortiz1}. Our results demonstrate that the universal
control achieved by the liquid state NMR QIPs enables efficient
simulation of some fermionic (and other particle statistics) systems,
providing relevant information about the particular phenomenon or
system of study \cite{note2}. In particular, we show how the spectrum
of the Fano-Anderson Hamiltonian can be determined.

The paper is organized in the following way: In Sec. \ref{standardmodel}
we introduce the conventional model of quantum computation and use it
to describe the physics of the liquid state NMR setting as a universal
quantum simulator.  In Sec.~\ref{simulation} we show quantum algorithms
for obtaining relevant physical properties of quantum systems
satisfying different particle statistics, by mapping their algebras of
operators into the spin-1/2 algebra (conventional model). In
Sec.~\ref{fanoanderson} we introduce the fermionic Fano-Anderson model,
and show how to simulate it in the liquid state NMR device. Its
experimental implementation as well as the results, and the conclusions
are described in Sec.~\ref{experimentalimplementation} and
Sec.~\ref{conclusions}, respectively.

\section{Quantum information processing with liquid state NMR methods}
\label{standardmodel}

In this section we introduce liquid state NMR quantum information
processing methods, emphasizing the fact that they can be
mathematically described in terms of Pauli (spin-1/2) operators
\cite{ni1}.  A more detailed description of such methods can be found
in \cite{NMR1}.

In the conventional model of quantum computation the fundamental unit
of information is the quantum bit or {\it qubit}. A qubit's pure
state, $\ket{\sf a}=a \ket{0} + b \ket{1}$ (with $a,b \in \mathbb{C}$
and $|a|^2 + |b|^2=1$), is a linear superposition of the logical
states $\ket{0}$ and $\ket{1}$, and can be represented by the state of
a two-level quantum system such as a spin-1/2. Similarly, a pure state
of a register of $N$ qubits is represented as $\ket{\psi}=\sum_{n=0}
^{2^N-1} a_n \ket{n}$, where $\ket{n}$ is a product of states of each
qubit in the logical basis, e.g., its binary representation
($\ket{0}\equiv \ket{00 \cdots0},
\ket{1}\equiv\ket{00\cdots01},\ket{2}\equiv\ket{00\cdots10}$, etc.),
and $\sum_{n=0}^{2^N-1} |a_n|^2=1$ ($a_n \in \mathbb{C}$). A quantum
register can also be in a probabilistic mixture of pure states, i.e.,
a mixed state, which is described by a density matrix $\rho=\sum_s p_s
\rho_s$, with $\rho_s=\ket{\psi_s}\bra{\psi_s}$ representing the state
of the register in the pure state $\ket{\psi_s}$, with probability
$p_s$. Every density operator can be written as a sum of products of the Pauli
spin-1/2 operators $\sigma_\alpha^j$ ($\alpha=x,y,z$, and
$j=[1,\cdots, N]$) and the identity operators $\id^j$ acting on the
$j$-th qubit of the register \cite{NMR1}.

\begin{figure}[hbt]
\begin{center}
\includegraphics[height=5.5cm]{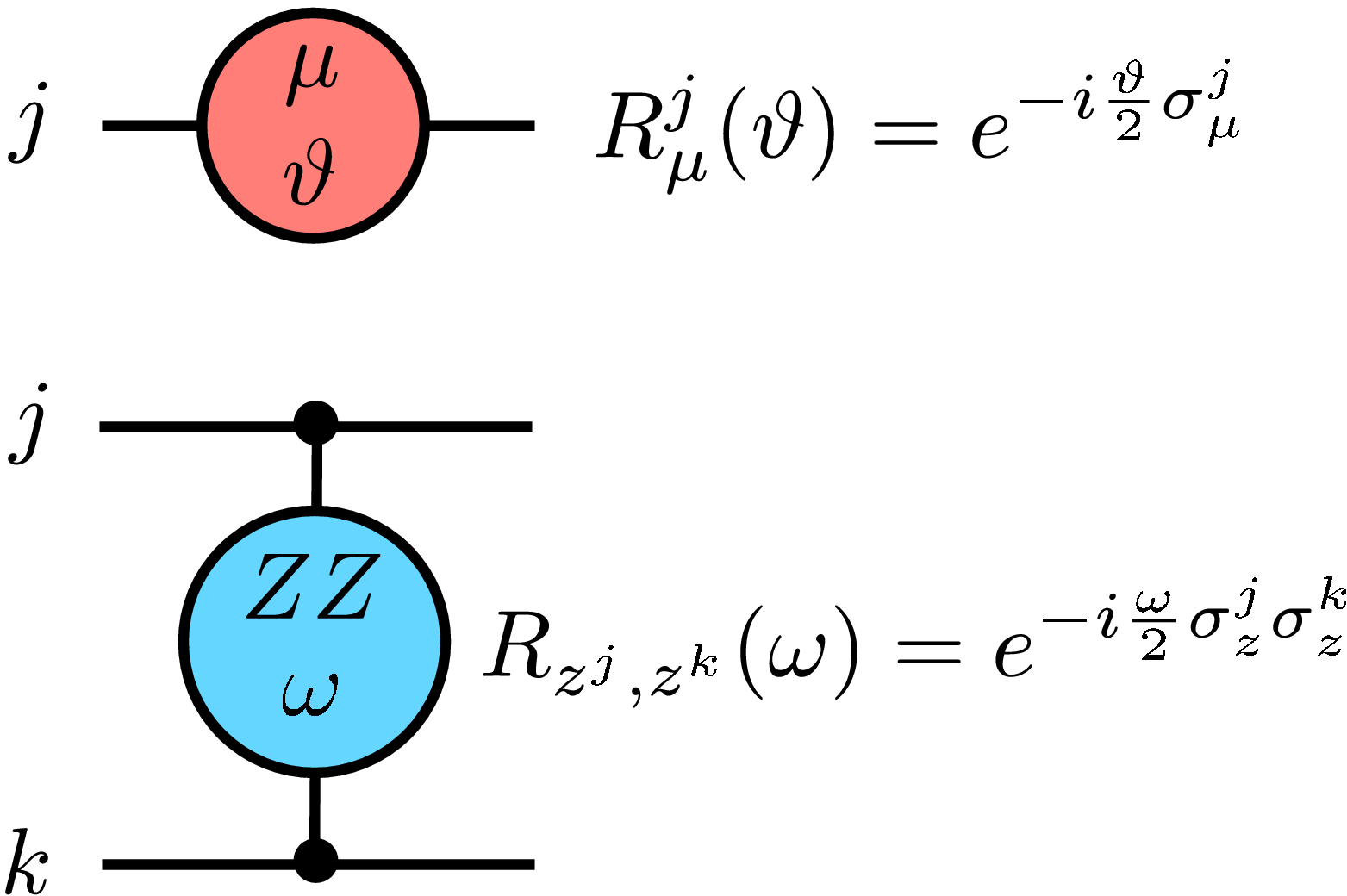}
\end{center}
\caption{Circuit representation of the elementary gates.  The top
picture indicates a single-qubit rotation while the bottom one
indicates the two-qubit Ising gate.  Any quantum algorithm can be
represented by a circuit composed of these elementary gates (see for
example Fig.~\ref{refocalisation})}.
\label{sm1}
\end{figure}

The Pauli operators can also be used to describe any unitary operation
acting on the state of the register. In particular, every
unitary operation can be decomposed in terms of single-qubit rotations
$R_\mu^j (\vartheta)= e^{-i\frac{\vartheta}{2} \sigma_\mu^j} =[ \cos
(\vartheta/2)\id^j - i \sin (\vartheta/2)\sigma_\mu^j]$, by an angle
$\vartheta$ around the $\mu$-axis, and two-qubit interactions such as the
\emph{Ising gate} $R_{z^j,z^k}(\omega) = e^{-i \frac{\omega}{2}
\sigma_z^j \sigma_z^k} = [\cos(\omega/2)\id^j \id^k -i \sin
(\omega/2)\sigma_z^j\sigma_z^k]$ \cite{ba1,vi1}, defining a
universal set of elementary gates. In Fig. \ref{sm1} we show the
quantum circuit representation of these basic operations.

Finally, in the conventional model of quantum computation the
measurement is assumed to be projective and is described by
projectors that can be expanded in terms of Pauli operators.

Liquid-state NMR methods allow us to physically implement a slightly
different version of the conventional model of quantum computation,
with respect to the initial state and the measurement process. In this
set-up the quantum register is represented by the average state of the
nuclear spin-1/2 of an ensemble of identical molecules. Since all
molecules are equivalent, in the following analysis we will first
consider only one of them. The spin state of each nucleus (qubit) of a
single molecule is manipulated using resonant radio-frequency magnetic
pulses (RF pulses).

\begin{figure}[hbt]
\begin{center}
\includegraphics[height=5.5cm]{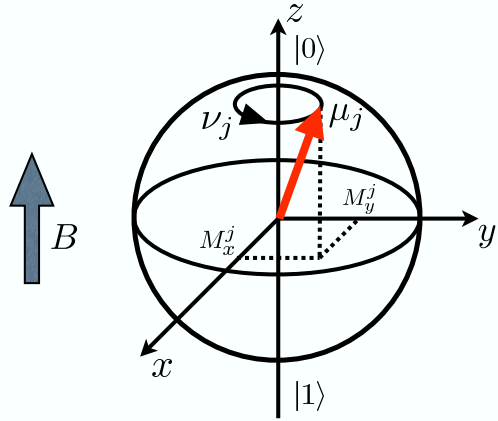}
\end{center}
\caption{Bloch's sphere representation of a single nuclear spin-1/2
precessing around the quantization axis determined by the external
magnetic field $B$. The precession frequency is given by $\nu_j = \mu_j
B$, with $\mu_j$ the magnetic moment of the $j$-th nucleus. Due to the
chemical environment, each nucleus precesses at its own Larmor
frequency $\nu_j$.} 
\label{larmor}
\end{figure}

The molecule is placed in a strong magnetic field $B(\hat{z}) \simeq
10$ Tesla, so that the spin of the $j$-th nucleus precesses at its
(Larmor) frequency $\nu_j$ (Fig.~\ref{larmor}).  In the frame rotating
with the $j$-th spin, its qubit state can then be rotated by sending
RF pulses in the $x$-$y$ plane at the resonant frequency $\nu_r \sim
\nu_j$. If the duration of this pulse is $\Delta t$, the corresponding
evolution operator in the rotating frame is \cite{NMR1}
\begin{equation}
U_j =  e^{-i H_j \Delta t} = e^{-i {\sf A} (\cos(\varphi) \sigma_x^j 
+ \sin(\varphi) \sigma_y^j) \Delta t} ,
\end{equation} 
where ${\sf A}$ is the amplitude of the RF-pulse and $\varphi$ is its
phase in the $x$-$y$ plane ($\hbar=1$). Then one can induce single spin
rotations \cite{nota1} along any axis in the $x$-$y$ plane by adjusting
$\Delta t$ and $\varphi$.

Single-qubit rotations around the $z$-axis can be implemented with no
experimental imperfection or physical duration simply by changing the
phase of the abstract rotating frame we are working with. We have then
to keep track of all these phase changes with respect to a reference
phase associated with the spectrometer. Nevertheless, these phase
tracking calculations are linear with respect to the number of pulses
and spins, and can be efficiently done on a classical
computer. Together with the rotations along the $x$- or $y$-axis, the
$z$-rotations can generate any single qubit rotation on the Bloch
sphere.

On the other hand, the spin-spin interactions present in the molecule
allow us to perform two-qubit gates and achieve universal control. To
first order in perturbation, this interaction (called the $J$-coupling), has
the form
\begin{equation}
\label{zzinteraction}
H_{j,k} = \frac{J_{jk}}{4} \sigma_z^j \sigma_z^k ,
\end{equation}
where $j,k$ denote the corresponding pair of qubits and $J_{jk}$ is
their coupling strength. Under typical NMR operating conditions, these
interaction terms are small enough to be neglected when performing
single-qubit rotations with RF pulses of short duration . Nevertheless,
between two pulses they are driving the evolution of the system. By
cleverly designing a pulse sequence, i.e., a succession of pulses and
free evolution periods, one can easily apply two-qubit gates on the
state of the system. Indeed, the so-called {\it refocusing techniques}'
principle consists of  performing an arbitrary Ising gate by flipping
one of the coupled spins ($\pi$-pulse), as shown in Fig.
\ref{refocalisation}. The interaction evolutions before and after the
refocusing pulse compensate leading to the effective evolution
\begin{equation}
U^{\sf eff}_{j,k} = e^{i \frac{\pi}{2} \sigma_x^j} e^{-i
\frac{J_{jk}}{4}\sigma_z^j \sigma_z^k \Delta t_2} e^{-i \sigma_x^j
\pi/2} e^{-i\frac{J_{jk}}{4} \sigma_z^j \sigma_z^k \Delta t_1} = e^{-i
\frac{\bar{\alpha}}{4} \sigma_z^j \sigma_z^k} ,
\end{equation}
where the effective coupling strength $\bar{\alpha}= J_{jk} (\Delta
t_1-\Delta t_2)$ is being determined by the difference between the
durations $\Delta t_1$ and $\Delta t_2$.

We have so far described a quantum register as consisting of nuclei of
a single molecule. However, liquid state NMR uses an ensemble of about
$10^{23}$ molecules in a solution maintained at room temperature
($\simeq 300K$). For typical values of the magnetic field, this
thermal state is extremely mixed. Clearly, this is not the usual state
in which we initialize a quantum computation since qubits are nearly
randomly mixed.  Nevertheless, known NMR methods \cite{NMR1} can be
used to prepare the so-called \emph{pseudo-pure state} ($\rho_{\sf
pp}$) \cite{psudop}
\begin{equation}
\label{pseudopure}
\rho_{{\sf pp}} = \frac{(1-\epsilon)}{2^N} \id + \epsilon\rho_{{\sf
pure}} ,
\end{equation} 
where $\rho_{{\sf pure}}$ is a density operator that describes a pure
state and $\epsilon$ is a small real constant (i.e., $\epsilon$ decays
exponentially with $N$).

Under the action of any unitary transformation $U$ this state evolves
as
\begin{equation}
\label{ppstate}
\rho_{\sf pp}^{\sf final} = U \rho_{\sf pp} U^\dagger =
\frac{(1-\epsilon)}{2^N}I + U \epsilon\rho_{{\sf pure}} U^\dagger .
\end{equation}
The first term in Eq. \ref{ppstate} did not change because the
identity operator is invariant under any unitary
transformation. Therefore, performing quantum computation on the
ensemble is equivalent to performing quantum computation over the initial
state represented only by $\rho_{{\sf pure}}$.  

After the quantum computation is performed, we measure the orthogonal
components of the sample polarization in the $x$-$y$ plane, $M_x = {\sf
Tr} (\rho_{\sf pp}^{\sf final} \sum_{i=1}^N \sigma_x^i)$, and $M_y = {\sf
Tr}(\rho_{\sf pp}^{\sf final} \sum_{i=1}^N \sigma_y^i)$. Note that the
invariant component of $\rho_{\sf pp}^{\sf final}$ does not contribute
to the signal since ${\sf Tr}(I\sigma_{x,y}^j)=0$. Since the
polarization of each single spin, $M_x^j= {\sf Tr} (\rho_{\sf pp}^{\sf
final} \sigma_x^j)$ and $M_y^j= {\sf Tr} (\rho_{\sf pp}^{\sf final}
\sigma_y^j)$, precesses at its own Larmor frequency $\nu_j$, a Fourier
transformation of the temporal recording (called FID, for Free Induction
Decay) of the total magnetization needs to be performed. By doing so,
we obtain the expectation value of the polarization of each spin
(averaged over all molecules in the sample).

Summarizing, a liquid state NMR setting allows us to initialize a
register of qubits in a pseudo-pure state, apply any unitary
transformation to this state by sending controlled RF pulses or by free
interaction periods, and measure the expectation value of some quantum
observables (i.e., the spin polarization). Hence, these systems can be
used as quantum information processors (QIPs).

\section{Simulation of Physical Systems}
\label{simulation}

Richard P. Feynman \cite{feynman} described a quantum computer as a
universal reversible device governed by the laws of quantum physics
and capable of exactly simulating any physical system.  Although he
analyzed the problem of simulating physics assuming that every finite
quantum mechanical system can be imitated exactly by another one (e.g.,
a set of qubits) \cite{note1}, he was unsure whether this statement
remained valid for the simulation of fermionic systems.

In this section we describe how to obtain information about physical
properties of any quantum many-body system (fermionic, bosonic,
anyonic, etc.)  by using a set of qubits (spin-1/2) controlled by NMR
techniques. A more complete description of these methods based on the
existence of one-to-one mappings between the algebras used to describe
the system to be simulated and the quantum computer
\cite{jordan1,batista1,batista2}, as well as indirect measurement
algorithms \cite{ortiz1}, can be found in previous works
\cite{ortiz1,somma1,somma2}.  

\begin{figure}[hbt]
\begin{center}
\includegraphics[height=5.5cm]{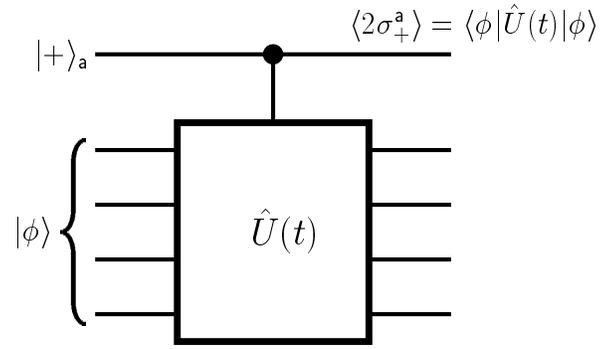}
\end{center}
\caption{Quantum network for the evaluation of the expectation value of
a unitary operator $\hat{U}(t)$. The filled circle denotes a controlled
operation (i.e., ${\sf U^{\ket{1}_a}}$ of Sec.
\ref{simulation}), such that $\hat{U}(t)$ is applied to the
system only if the ancilla qubit is in the state $\ket{1}_{\sf a}$.  }
\label{qa1}
\end{figure}

In this work we are interested in the measurement of correlation
functions of the form
\begin{equation} 
\label{correl1}
G(t)=\bra{\phi} \hat{U}(t) \ket{\phi },
\end{equation} 
where $\hat{U}(t)$ is any time (or other continuous parameter)
dependent unitary operator, using indirect measurement techniques
\cite{ortiz1}. In addition to the qubits used to {\it represent} the
physical system to be simulated (i.e., the system of qubits), an extra
qubit called {\it ancilla} is required (Fig. \ref{qa1}). This qubit
will be used as a probe to scan the properties of the system of qubits. 
It has to be initialized in the superposition state $\ket{+}_{\sf
a}=\frac{\ket{0}_{\sf a}+\ket{1}_{\sf a}} {\sqrt{2}}$ by applying the
Hadamard gate \cite{had} to the polarized state $\ket{0}_{\sf a}$. 
Then, it interacts with the system of qubits, initially in the state
$\ket{\phi}$, through a controlled unitary operation ${\sf
U^{\ket{1}_{\sf a}} }= \ket{0}_{\sf a} \langle 0 | \otimes \id +
\ket{1}_{\sf a} \langle 1 | \otimes \hat{U}(t)$. After this
interaction, we can show \cite{ortiz1} that $G(t) = \langle 2
\sigma_+^{\sf a}\rangle = \langle \sigma_x^{\sf a} + i \sigma_y^{\sf a}
\rangle$; that means we get the desired result by measuring the
expectation values of the ancilla qubit observables $\sigma_x^{\sf a}$,
and $\sigma_y^{\sf a}$.

Using the same techniques we can determine the spectrum of an
observable $\hat{Q}$ when choosing $\hat{U}(t)= e^{-i \hat{Q} t}$.
Figure \ref{qa2} depicts this algorithm  \cite{somma1}.  Since the
initial state can always be written as a linear combination of
eigenstates of $\hat{Q}$, that is, $\ket{\phi} = \sum\limits_{n}
\gamma_n \ket{\psi_n}$, with $\ket{\psi_n}$ the eigenstates of
$\hat{Q}$ having eigenvalues $\lambda_n$, and $\gamma_n$ complex
coefficients, a measurement on the polarization of the ancilla qubit
gives $ \langle 2 \sigma_+^{\sf a}(t) \rangle = \sum\limits_{n} |
\gamma_n |^2 e^{-i \lambda_n t}$. Having the time-dependent function 
$S(t)=\langle 2 \sigma_+^{\sf a}(t) \rangle$ for a discrete set of
values $t_i$, the eigenvalues $\lambda_n$ can in principle be obtained
by performing a discrete Fourier transform (DFT) \cite{somma1}. Note
that the determination of each single value $S(t_i)$ requires a
different experiment.

The eigenvalues $\lambda_n$ denote the spectrum of a system Hamiltonian
$H$ when replacing $\hat{Q} \rightarrow H$. In this case, the operation
${\sf U^{\ket{1}_a}}$ can be efficiently
implemented~\cite{ortiz1,somma1,somma2}. However, methods for finding
an initial state with an overlap $\gamma_n$ that does not vanish
exponentially with increasing system size, are in general not known.
This issue arises, for example, when trying to obtain the spectrum of
the two-dimensional Hubbard model approaching the thermodynamic
limit~\cite{somma1,somma2}.

Nevertheless, the same basic procedure can be used when interested in
obtaining dynamical correlation functions of the form $G(t)=\bra{\phi}
T^\dagger A_i T B_j \ket{\phi}$ (i.e., $\hat{U}(t) = T^\dagger A_i T
B_j$ in Eq. \ref{correl1}), where $T=e^{-i H t}$ is the time evolution
operator of a time-independent Hamiltonian $H$, and $A_i$, $B_j$
are unitary operators.  In Fig.  \ref{qa3} we show the circuit for an
algorithm capable of obtaining these correlation functions after some
simplifications \cite{somma1}.  The evolution has three different
steps: First, we perform a controlled operation ${\sf B^{\ket{1}_a} } =
\ket{0}_{\sf a} \langle 0 | \otimes \id + \ket{1}_{\sf a} \langle 1 |
\otimes B_j$.  Second, we perform the $T$ operation on the system, and
third, a controlled operation ${\sf A^{\ket{0}_a}} =
\ket{0}_{\sf a} \langle 0 | \otimes A^\dagger_i + \ket{1}_{\sf a}
\langle 1 | \otimes \id$. Spatial correlation functions can also be
obtained when replacing the operator $T$ by the space translation
operator. Again, this algorithm can be performed efficiently whenever
the initial state $\ket{\phi}$ can be prepared efficiently.

The algorithm described above can be easily implemented with
liquid-state NMR methods, since the result of the simulation is encoded
in the expectation values of single qubit observables. So far, the
algorithm applies only to the simulation of systems described in terms
of Pauli operators, such as spin-1/2 systems. However, other systems with
different particle statistics can also be simulated with these
algorithms after mapping their operator algebras onto the Pauli
spin-1/2 algebra \cite{jordan1,batista1,batista2}. In the next section
we introduce the Fano-Anderson model, a simple fermionic system, and
show how to simulate it on a liquid-state NMR QIP using these methods.

\section{The Fano-Anderson model}
\label{fanoanderson}

The quantum simulation of the one-dimensional fermionic
Fano-Anderson model provides a starting point for simulations of
quantum systems with different kinds of particle statistics.

\begin{figure}[hbt]
\begin{center}
\includegraphics[height=7.5cm]{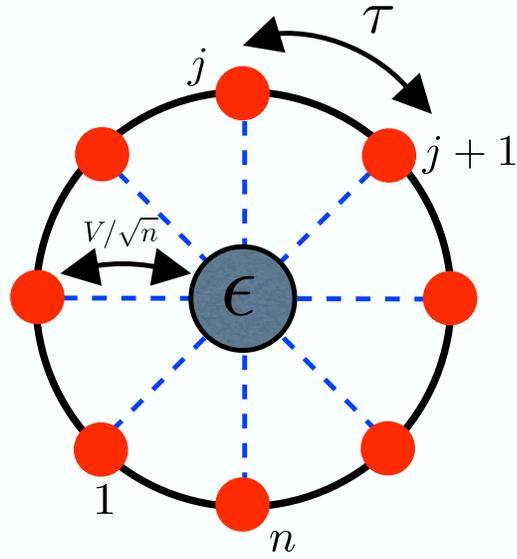}
\end{center}
\caption{Fermionic Fano-Anderson model. Fermions can hop between
nearest-neighbor sites (exterior circles) and between a site and the
impurity (centered circle), with hopping matrix elements $\tau$ and
$V/\sqrt{n}$, respectively. The energy of a fermion in the impurity is
$\epsilon$.}
\label{fa1}
\end{figure}

The one-dimensional fermionic Fano-Anderson model consists of an
$n$-sites  ring with an impurity in the center (see Fig. \ref{fa1}), 
where spinless fermions can hop between nearest-neighbors sites with
hopping matrix element (overlap integral) $\tau$, or between a site and
the impurity with matrix element $V/\sqrt{n}$. Taking the
single-particle energy of a fermion in the impurity to be $\epsilon$,
and considering the translational invariance of the system, the
Fano-Anderson Hamiltonian can be written in the wave vector
representation as \cite{ortiz1}
\begin{equation}
\label{Hamilt2}
H=\sum_{l=0}^{n-1} \varepsilon_{k_l} c^{\dagger}_{k_l}c_{k_l}^{\;}+
\epsilon b^{\dagger}b + V(c^{\dagger}_{k_0}b+b^{\dagger}c_{k_0}^{\;}),
\end{equation}
where the fermionic operators $c^{\dagger}_{k_l}$ and $b^\dagger$
($c_{k_l}^{\;}$ and $b$) create (destroy) a spinless fermion in the
conduction mode $k_l$ and in the impurity, respectively.  Here, the
wave vectors are $k_l=\frac{2\pi l}{n}$ ($l=[0,..,n-1]$) and the
energies per mode are $\varepsilon_{k_l} = -2 \tau \cos k_l$.

In this form, the Hamiltonian in Eq.~ \ref{Hamilt2} is almost diagonal
and can be exactly solved: There are no interactions between electrons
in different modes $k_l$, except for the mode $k_0$, which interacts
with the impurity. Therefore, the relevant physics comes from this
latter interaction, and its spectrum can be exactly obtained by
diagonalizing a $2 \times 2$ Hermitian matrix, regardless of $n$ and
the number of fermions in the ring $N_e$. Nevertheless, its simulation
in a liquid-state NMR QIP is the first step in quantum simulations of
quantum many-body problems. 

\begin{figure}[hbt]
\begin{center}
\includegraphics[height=10cm]{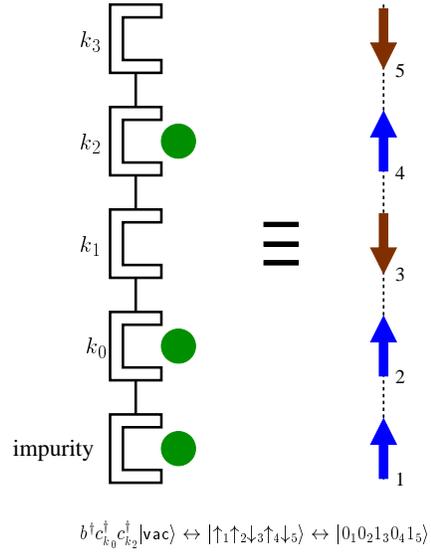}
\end{center}
\caption{Mapping of the fermionic product state $c^\dagger_1
c^\dagger_2 c^\dagger_4 \ket{{\sf vac}}$, with $\ket{{\sf vac}}$ the
no-fermion or vacuum state, into the spin-1/2 and the standard quantum
computation languages, using the Jordan-Wigner transformation. A filled
circle denotes a site occupied by a spinless fermion, which maps into
the state $\ket{\uparrow}$ in the  spin 1/2 algebra.}
\label{ps1}
\end{figure}

In order to use the algorithms presented in Sec.~\ref{simulation}, and
to successfully simulate this system in an NMR QIP, we first need to
map the fermionic operators onto the spin-1/2 (Pauli) operators. This
is done by use of the following Jordan-Wigner
transformation~\cite{jordan1}
\begin{equation}
\label{jwmap2}
\matrix {b = \sigma_-^1 & b^\dagger = \sigma_+^1 \cr c^{\;}_{k_0} = -
\sigma_z^1 \sigma_-^2 & c^{\dagger}_{k_0} = - \sigma_z^1 \sigma_+^2
\cr \vdots & \vdots \cr 
c^{\;}_{k_{n-1}} = \left ( \prod_{j=1}^n -\sigma_z^j \right )
\sigma_-^{n+1}  & \qquad c^{\dagger}_{k_{n-1}} = \left ( \prod_{j=1}^n
-\sigma_z^j \right )\sigma_+^{n+1} . }
\end{equation}
In this language, a logical state $\ket{0_j}$ (with $\ket{0}\equiv
\ket{\uparrow}$ in the usual spin-1/2 notation) corresponds to having a
spinless fermion in either the impurity, if $j=1$, or in the mode
$k_{j-2}$, otherwise. The fermionic vacuum state $\ket{{\sf vac}}$
(i.e., the state with no fermions) maps onto $\ket{{\sf \widehat{vac}}}
= \ket{1_1 1_2 \cdots 1_{n+1}}$ ($\equiv\ket{ \downarrow_1
\downarrow_2  \cdots \downarrow_{n+1}}$). As an example, Fig.~\ref{ps1}
shows the mapping of a particular fermionic state for $n=4$.

Some dynamical properties of this model can be obtained using the
quantum algorithms described in Sec.~\ref{simulation}.  Here,
we are primarily interested in obtaining the probability amplitude of
having a fermion in mode $k_0$ at time $t$, if initially ($t=0$)
the quantum state is the Fermi sea state with $N_e$ fermions; that is,
$\ket{\sf FS}=\prod\limits_{l=0}^{N_e-1} c^\dagger_{k_l} \ket{{\sf
vac}}$. This probability is given by the modulus square of the
following dynamical correlation function: 
\begin{equation}
\label{correlation}
G(t)= \bra{\sf FS} b(t) b^\dagger(0) \ket{\sf FS} \ ,
\end{equation}
where $b(t) = T^\dagger b(0) T$, $T=e^{-i Ht}$ is the time evolution
operator, and $b^\dagger(0)=b^\dagger$. Basically, $G(t)$ is the
overlap between the quantum state $ b^\dagger(0) \ket{\sf FS}$, which
does not evolve, and the state $b^\dagger(t) \ket{\sf FS} $, which does
not vanish unless the evolved state $T\ket{\sf FS}$ already contains a
fermion in the impurity site ($(b^\dagger(t))^2 =(b^\dagger(0))^2= 0$).
In terms of spin-1/2 operators (see Eq. \ref{jwmap2}), this correlation
function reduces to a two-qubit problem \cite{ortiz1}:
\begin{equation}
\label{correl2}
G(t)= \bra{\phi} \bar{T}^\dagger \sigma_-^1 \bar{T} \sigma_+^1
\ket{\phi} \ ,
\end{equation}
where $\bar{T} = e^{-i \bar{H} t}$ is an evolution operator arising
from the interaction terms in Eq. \ref{Hamilt2}, with
\begin{equation}
\label{Hamilt4}
\bar{H} = \frac{\epsilon}{2} \sigma_z^1 + \frac{\varepsilon_{k_0}}{2}
\sigma_z^2 + \frac{V}{2} (\sigma_x^1 \sigma_x^2 + \sigma_y^1
\sigma_y^2) \ ,
\end{equation}
and $\ket{\phi} = \ket{1_1 0_2}$ in the logical basis (i.e., the
initial state with one fermion in the $k_0$ mode).

\begin{figure}[hbt]
\begin{center}
\includegraphics[height=7.5cm]{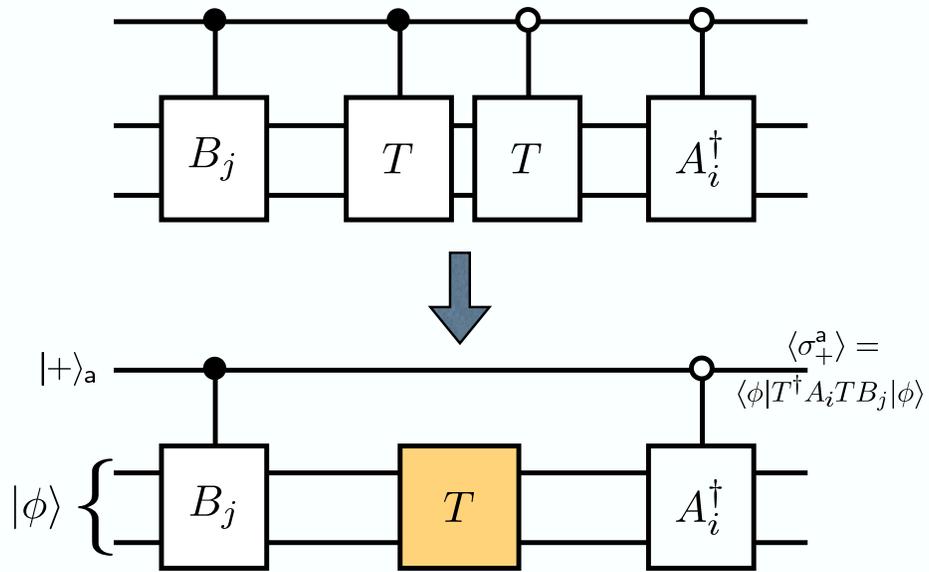}
\end{center}
\caption{Quantum network for the evaluation of the correlation function
$G(t)=\langle \phi \ket{T^\dagger A_i T B_j |\phi}$. The filled (empty)
circle denotes an operation controlled in the state $\ket{1}_{\sf a}$  
($\ket{0}_{\sf a}$) of the ancilla qubit.}
\label{qa3}
\end{figure}

\begin{figure}[hbt]
\begin{center}
\includegraphics[height=10.cm]{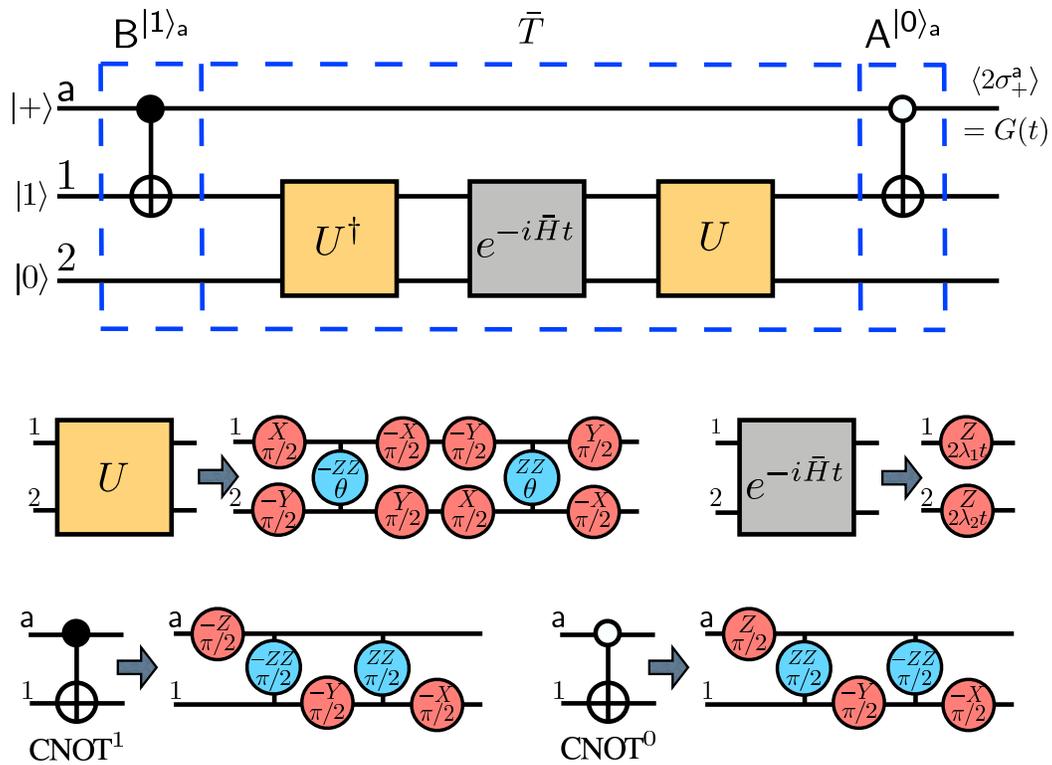}
\end{center}
\caption{Quantum circuit for the evaluation of $G(t)$ (Eq.
\ref{correlation}) in terms of elementary gates directly
implementable with liquid-state NMR methods.  }
\label{fa2}
\end{figure}

In order to use the quantum circuit shown in Fig.~\ref{qa3}, all 
operators in Eq. \ref{correl2} must be unitary. Using the symmetries of
$H$, such as the global $\pi/2$ $z$-rotation that maps $(\sigma_x^j ,
\sigma_y^j ) \rightarrow (\sigma_y^j , -\sigma_x^j )$, leaving the state
$\ket{\phi}$ invariant (up to a phase factor), we obtain
$\langle \phi | \bar{T}^\dagger \sigma_x^1 \bar{T} \sigma_y^1 \ket{\phi}=
\langle \phi | \bar{T}^\dagger \sigma_y^1 \bar{T} \sigma_x^1 \ket{\phi}=0$
and 
$\langle \phi | \bar{T}^\dagger \sigma_x^1 \bar{T} \sigma_x^1 \ket{\phi}=
\langle \phi | \bar{T}^\dagger \sigma_y^1 \bar{T} \sigma_y^1 \ket{\phi}$.
Then, Eq. \ref{correl2} can be written in terms of unitary operators as
\begin{equation}
\label{correl3}
G(t) = \bra{\phi}  e^{i \bar{H} t} \sigma_x^1 
e^{-i \bar{H} t}  \sigma_x^1 \ket{\phi}.
\end{equation}
Figure \ref{fa2} shows the quantum circuit used to obtain $G(t)$.  It
is derived from Fig.~\ref{qa3} by making the following identifications:
$T \rightarrow e^{-i \bar{H} t}$, $ A_i \rightarrow \sigma_x^1$, and $
B_j \rightarrow \sigma_x^1$. As we can see, the corresponding
controlled operations ${\sf A^{\ket{0}_a}}$ and ${\sf B^{\ket{1}_a}}$
transform into the well-known controlled-not (${\sf CNOT}$) gates.  All
the unitary operations appearing in Fig. \ref{fa2} were decomposed into
elementary NMR gates (single qubit rotations and Ising interactions).
In particular, the decomposition of $e^{-i \bar{H} t}$ can be found in
Ref. \cite{ortiz1}. We obtain
\begin{equation}
\label{decomp}
e^{-i \bar{H} t} = U e^{-i \lambda_1 \sigma_z^1 t} e^{-i \lambda_2
\sigma_z^2 t} U^\dagger \ ,
\end{equation}
where $\lambda_{1(2)} = \frac{1}{2} (E \mp \sqrt{\Delta^2 +V^2})$, with
$E= \frac{\epsilon + \varepsilon_{k_0}}{2}$, and  $\Delta=
\frac{\epsilon - \varepsilon_{k_0}}{2}$. The unitary operator $U$ is
decomposed as (Fig.~\ref{fa2})
\begin{equation}
\label{Adecomp}
U = e^{i \frac{\pi}{4} \sigma^2_x} e^{-i \frac{\pi}{4} \sigma^1_y}
e^{-i \frac{\theta}{2} \sigma^1_z \sigma^2_z} e^{i
\frac{\pi}{4}\sigma^1_y} e^{i \frac{\pi}{4} \sigma^1_x} e^{-i
\frac{\pi}{4}\sigma^2_x} e^{-i \frac{\pi}{4} \sigma^2_y} e^{i
\frac{\theta}{2}\sigma^1_z \sigma^2_z} e^{-i \frac{\pi}{4} \sigma^1_x}
e^{i \frac{\pi}{4} \sigma^2_y} ,
\end{equation}
with the parameter $\theta$ satisfying $\cos \theta =
1/\sqrt{1+\delta^2}$, and $\delta=(\Delta+\sqrt{\Delta^2+V^2})/V$. 

The {\sf CNOT} gates ${\sf A^{\ket{0}_a} }$ and ${\sf B^{\ket{1}_a} }$
can also be decomposed into elementary gates, obtaining ${\sf
A^{\ket{0}_a} }= \ket{0}_{\sf a} \langle 0 | \otimes \sigma_x^1 +
\ket{1}_{\sf a} \langle 1 | \otimes \id= e^{i \frac{\pi}{4} \sigma_x^1}
e^{i \frac{\pi}{4} \sigma_z^1\sigma_z^{\sf a}} e^{-i \frac{\pi}{4}
\sigma_y^1} e^{-i \frac{\pi}{4} \sigma_z^1\sigma_z^{\sf a}} e^{-i
\frac{\pi}{4} \sigma_z^1}$ and ${\sf B^{\ket{1}_a} }= \ket{0}_{\sf a}
\langle 0 | \otimes \id + \ket{1}_{\sf a} \langle 1 | \otimes
\sigma_x^1= e^{i \frac{\pi}{4} \sigma_x^1} e^{-i \frac{\pi}{4}
\sigma_z^1\sigma_z^{\sf a}} e^{-i \frac{\pi}{4} \sigma_y^1} e^{i
\frac{\pi}{4} \sigma_z^1\sigma_z^{\sf a}} e^{-i \frac{\pi}{4}
\sigma_z^1}$ (up to a phase factor).  In this way, we can proceed to
simulate the circuit of Fig.~\ref{fa2} and obtain $G(t)$ in an NMR QIP
by applying the appropriate RF pulses (Sec.~\ref{standardmodel}).  Only
three qubits are required for its simulation (Fig.  \ref{fa2}): The
ancilla qubit ${\sf a}$, one qubit representing the impurity site
(qubit-1), and one qubit representing the $k_0$ mode (qubit-2).

\begin{figure}[hbt]
\begin{center}
\includegraphics[height=5.5cm]{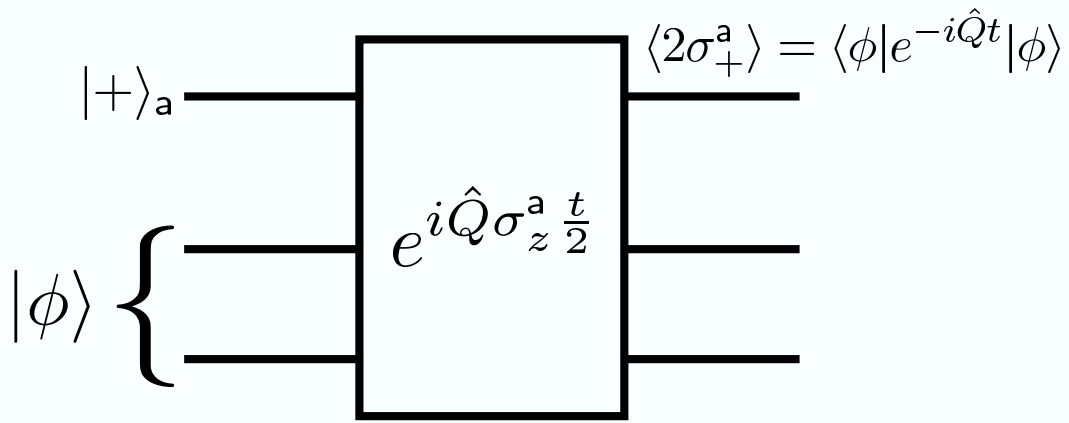}
\end{center}
\caption{Quantum network for the evaluation of the spectrum of an
observable $\hat{Q}$. }
\label{qa2}
\end{figure}

We are also interested in obtaining the spectrum of the Hamiltonian $H$
of Eq.~\ref{Hamilt2}. For this purpose we used the algorithm shown in
Fig.~\ref{qa2}, replacing $\hat{Q} \rightarrow H$.  In particular, when
$n=1$ (one site plus the impurity), Eq.~\ref{Hamilt2} reduces to $H=
\frac{\epsilon + \varepsilon_{k_0}}{2} +\bar{H}$, with $\bar{H}$
defined in Eq.~\ref{Hamilt4} in terms of Pauli operators. In this case,
the two eigenvalues $\lambda_i$ ($i=1,2$) of the one-particle subspace
can be extracted from the correlation function (Sec.~\ref{simulation})
\begin{equation}
\label{fanospectrum}
S(t)= \bra{\phi}  e^{-i H t} \ket{\phi} =  e^{-i (\epsilon +
\varepsilon_{k_0}) t} \bra{\phi} e^{-i \bar{H} t} \ket{\phi} ,
\end{equation}
which is equal to the polarization of the ancilla qubit after the
algorithm of Fig.~\ref{qa2} is performed. Since $\ket{\phi} =\ket{1_1
0_2}= \ket{\downarrow_1 \uparrow_2}$ is not an eigenstate of $H$, it
has a non-zero overlap with the two one-particle eigenstates, called
$\ket{{\sf 1P}_i}$ (see Appendix \ref{app1}).

\begin{figure}[hbt]
\begin{center}
\includegraphics[height=7.cm]{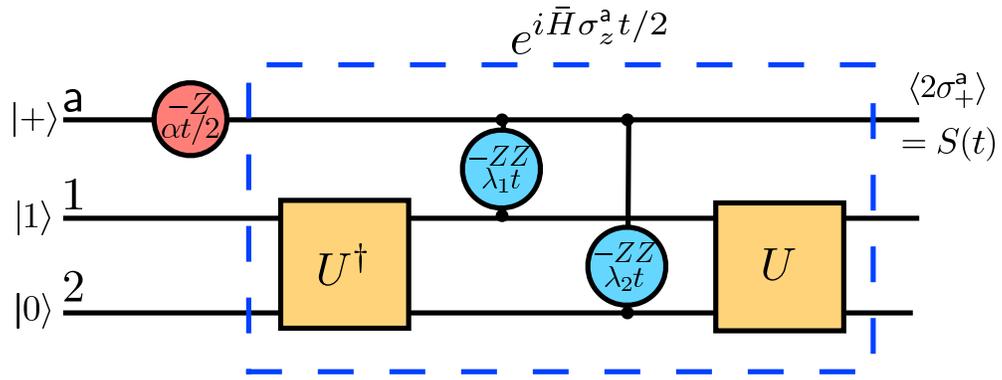}
\end{center}
\caption{Quantum circuit for the evaluation of $S(t)$ (Eq.
\ref{fanospectrum}). The parameters $\lambda_1 \mbox{ and }\lambda_2$
are defined in Sec. \ref{fanoanderson}, and $\alpha=\frac{\epsilon
+  \varepsilon_{k_0}}{2}$. The decomposition of the operator $U$ in NMR
gates can be found in Fig.~\ref{fa2}.}
\label{fa3}
\end{figure}

Again, the operator $e^{i H \sigma_z^{\sf a} t/2}$ (Fig.~\ref{qa2})
needs to be decomposed into elementary gates for its implementation in
an NMR QIP.  Noticing that $[ \sigma_z^{\sf a} , H ]= [ \sigma_z^{\sf
a} , U ]=0$, we obtain 
\begin{equation}
e^{i H \sigma_z^{\sf a} t/2} = U e^{i \lambda_1 \sigma_z^1
\sigma_z^{\sf a} t/2} e^{i \lambda_2 \sigma_z^2 \sigma_z^{\sf a} t/2}
U^\dagger e^{i (\epsilon + \varepsilon_{k_0})
\sigma_z^{\sf a} t/2},
\end{equation}
where the unitary operator $U$ is decomposed as in Eq. \ref{Adecomp}.
Figure \ref{fa3} shows the corresponding circuit in terms of elementary
gates. Again, qubits 1 and 2
represent the impurity site and the $k_0$ mode, respectively.
${\sf a}$ denotes the ancilla qubit.  Since the
idea is to perform a DFT on the results obtained from the measurement
(see Appendix \ref{app1}), we need to apply this circuit for several
values of $t$ (Sec.  \ref{simulation}).

\section{Experimental implementation}
\label{experimentalimplementation}
\subsection{Experimental protocol}
\label{protocol}

\begin{figure}
\includegraphics{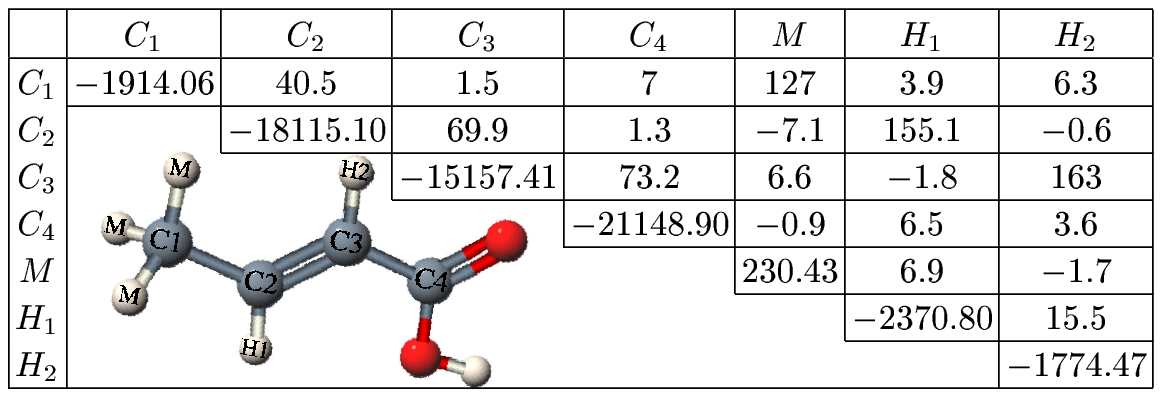}
\caption[Trans-crotonic acid molecule.]{The trans-crotonic acid
molecule is a seven-qubit register: The methyl group is used as a single
spin 1/2 \cite{refei2} and four $^{13}\text{C}$. The table shows in hertz
the values of the chemical shifts (on the main diagonal) and the
$J$-couplings (off-diagonal) between every pair of nuclei (qubits). }
\label{croto}
\end{figure}

For the experimental simulation of the fermionic Fano-Anderson model,
we used an NMR QIP based on a solution of trans-crotonic acid and
methanol dissolved in acetone. This setting has been described in Ref.
\cite{refei2}. Once the state of the 3 equivalent protons in the methyl
group of the trans-crotonic acid molecule is projected onto the spin-1/2
subspace \cite{refei2}, this molecule can be used as a seven-qubit
register (see Fig.  \ref{croto}). Methanol is used to perform RF-power
selection and accurately calibrate the RF pulses.

Two important characteristics of a molecule used for an NMR QIP are:
(i) the accuracy of the control and (ii) the number of elementary gates
we can perform within the relevant decoherence time of the system.  The
accuracy of control in trans-crotonic acid has been determined in Ref.
\cite{refei3}, using an error-correcting code as a benchmark.  The
current experiment can be considered as another exploration of the
accuracy of control, in this case examining how well we can implement
the necessary evolutions when simulating quantum systems with NMR
techniques.

In liquid-state NMR the main source of decoherence is the relaxation of
the transversal polarization of the sample due to the loss of coherence
between molecules. In our setting, the relevant times of this process,
called $T_2^*$, are in the range from several hundreds of milliseconds
to more than one second, for the different nuclei. These times fix the
maximum number of elementary gates that can be applied to the quantum
register without lossing coherence. Indeed, a lower bound of the pulse
duration to induce a rotation on a single qubit is determined by the
difference between the resonant frequencies of the spin to be rotated
and  the others (its chemical shift). A very short pulse having a wide
excitation profile in the frequency domain affects several spins at
the same time if their chemical shifts are small. On the other hand,
the duration of the Ising gate (two-qubit gate) depends directly on the
strength of the $J$-coupling constants $J_{jk}$.  In our setting the
chemical shifts values impose pulse durations of the order of 1 ms, and
the $J$-couplings impose interaction periods of the order of 10 ms,
restricting the pulse sequences to a maximum of approximately 1000
single-qubit rotations and 100 two-qubit (Ising) gates. 

Designing a pulse sequence to implement exactly the desired unitary
transformation would require very long refocusing schemes to cancel out
all the unwanted naturally occurring $J$-couplings. Then, the overall
duration of the pulse sequence increases and decoherence effects could
destroy our signal. Therefore, we need to find the best trade-off
between the ideal \cite{idealm} accuracy of the pulse sequence and its
duration, and neglect small couplings. For this purpose, we used an
efficient pulse sequence compiler to perform the phase tracking
calculations and to numerically optimize the delays between pulses, in
order to minimize the error that we introduce into the quantum
computation by neglecting small couplings.  

We now describe the parts of the pulse sequence corresponding
to the three basic steps of the quantum simulation.  

\paragraph{Pseudo-pure state preparation:} 
Initially, the state of the nuclei of the trans-crotonic acid molecules
in solution is given by the thermal distribution (Sec.
\ref{standardmodel}).  Using the methods described in Ref.
\cite{refei2} we have prepared the labeled pseudo-pure state (lpp)
$\rho_{\sf lpp} =\textbf{1}^{\text{C}_4}\textbf{1}^{\text{C}_3}
\textbf{1}^{\text{C}_2}\sigma_z^{\text{C}_1}\textbf{1}^{\text{M}}
\textbf{1}^{\text{H}_2}\textbf{1}^{\text{H}_1}$, where $\textbf{1}=
I-\sigma_z$ (i.e., $\textbf{1}=\ket{1}\langle 1|$) and
$\textbf{0}=I+\sigma_z$ (i.e., $\textbf{0}=\ket{0}\langle 0|$).  As we
will see, the state $\rho_{\sf lpp}$, having the spin of $\text{C}_1$
in the $\sigma_z$ state, is a good initial state for our purposes.

\paragraph{Initialization:} 

As mentioned in Sec. \ref{fanoanderson}, we need only 3 qubits to
simulate the Fano-Anderson model. These qubits must be well coupled to
each other to decrease the duration of the corresponding Ising gates we
apply to them. We have chosen the spin-1/2 nucleus $\text{C}_1$ to
represent qubit-1 (i.e., the impurity) and the spin-1/2 nucleus $M$ to
represent qubit-2 (i.e., the $k_0$ mode). On the other hand, we have
chosen the spin-1/2 nucleus $\text{C}_2$ to be the ancilla qubit ${\sf
a}$, to take advantage of its strong coupling with the spin-1/2 nucleus
$\text{C}_1$ (qubit-1). Since the rest of the spins
($\text{C}_4,\text{C}_3,\text{H}_2,\text{H}_1$) in the molecule remain
in the state $\textbf{1}$ or $\textbf{0}$ during the whole duration of
the experiment, we need to consider only the spins $\text{C}_2 \otimes
\text{C}_1 \otimes \text{M}$ with the above identification.

\begin{figure}[hbt]
\begin{center}
\includegraphics[height=5.cm]{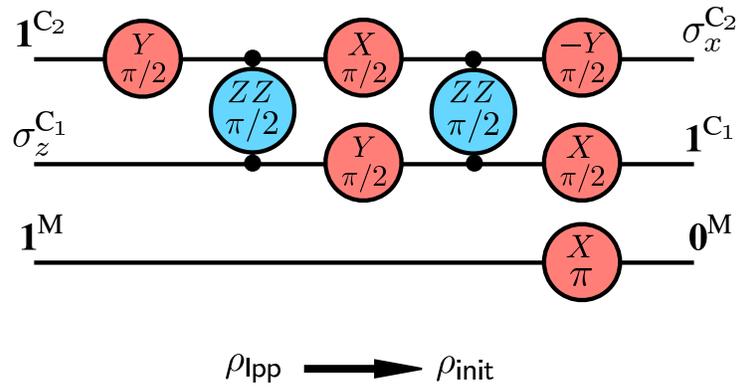}
\end{center}
\caption{Initialization pulse sequence used to transform the initial
labeled pseudo-pure state $\rho_{\sf lpp}=\textbf{1}^{\text{C}_2}
\sigma_z^{\text{C}_1} \textbf{1}^{\text{M}}$ into the state $\rho_{{\sf
init}} = \sigma_x^{\text{C}_2} \textbf{1}^{\text{C}_1}
\textbf{0}^{\text{M}}$. The sequence transfers the polarization from C$_1$ to
C$_2$ and flips the spin of the methyl group M. We have chosen  the spin-1/2
nuclei $\text{C}_2$, $\text{C}_1$, and $\text{M}$ to represent the ancilla,
qubit-1 (i.e., the impurity), and qubit-2 (i.e., the $k_0$-mode),
respectively.}
\label{ei1}
\end{figure}

The initial state $ \ket{+}_{\sf a} \otimes \ket{1_1 0_2}$
(Sec.~\ref{fanoanderson}) can be written as $\rho_{\sf init}' = \frac{1}{2}[
(I^{\sf a}+\sigma_x^{\sf a})\textbf{1}^1 \textbf{0}^2]$ in terms of
Pauli operators.  The ancilla qubit is only a \emph{control qubit} and
its state (i.e., its reduced density matrix) becomes correlated with
the rest of the qubits. Since the identity part is
not observable, we considered $\rho_{{\sf init}} = \sigma_x^{\sf
a}\textbf{1}^1 \textbf{0}^2$ instead of $\rho_{{\sf init}}'$ as the
initial state.  Its preparation was done by applying a sequence of
elementary gates to $\rho_{{\sf lpp}}=\textbf{1}^{\sf
a}\sigma^1_z\textbf{1}^2$, as shown in Fig.~\ref{ei1}.

\begin{figure}[hbt]
\begin{center}
\includegraphics[height=7.cm]{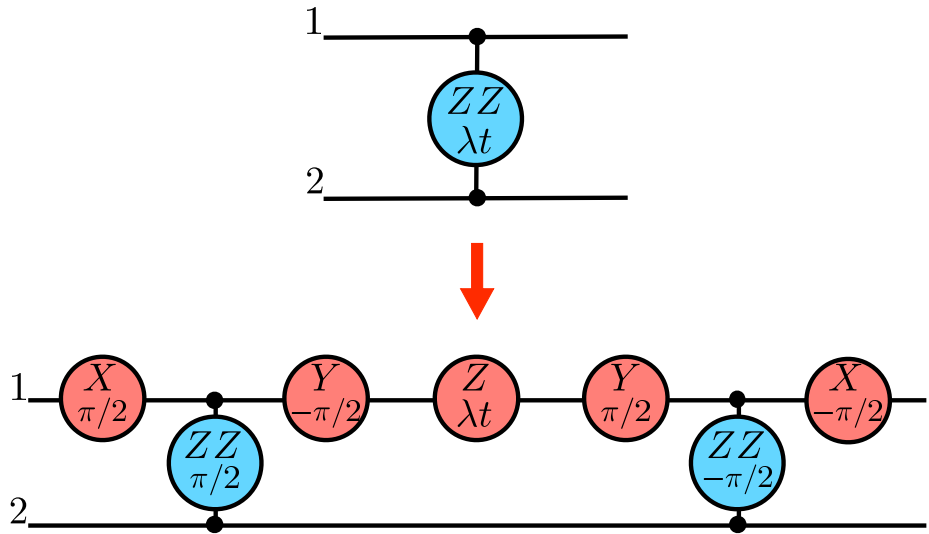}
\end{center}
\caption{Modification of a two-qubit gate with a coupling strength
depending on a parameter $t$. The variable interaction period is
translated into fixed interaction periods and a single-qubit rotation
with variable angle about the $z$-axis. Using this trick, the duration
of the physical pulse sequence does not depend on the parameter $t$
representing the time of the simulation.}
\label{ei2}
\end{figure}

\paragraph{Evolution pulse sequence:} 
As shown in Fig. \ref{fa3}, the pulse sequence used for obtaining
$S(t)$ (Eq.  \ref{fanospectrum}) requires Ising gates with a coupling
strength depending on $t$. The refocusing schemes are then optimized
differently and the results for different values of $t$ cannot be
directly comparable.  To avoid this problem we have replaced the two
Ising gates by an equivalent sequence of elementary gates, where the
dependence on the simulation parameter $t$ is transferred into the
angle of a single-qubit rotation along the $z$-axis (Fig. \ref{ei2}).
This {\it virtual} rotation is implemented through a phase tracking, as
mentioned in Sec. \ref{standardmodel}. Thus, the only difference
between the pulse sequence used to measure $S(t)$ for different
simulation times $t_i$ is a phase calculation that introduces no extra
optimization or experimental error.  

\paragraph{Measurement:} 
The result of the algorithm is encoded in the polarization of the
ancilla qubit $\langle 2 \sigma_+^{\sf a} \rangle = \langle
\sigma_x^{\sf a}\rangle +i \langle \sigma_y^{\sf a} \rangle $
(Sec.~\ref{simulation}), which is directly proportional to the
polarization of $\text{C}_2$ over the sample.  This component precesses
at the $\text{C}_2$ Larmor frequency $\nu_{\text{C}_2}$.  To measure
it, we have to perform a Fourier transformation on the measured FID and
integrate only the peak located at $\nu_{\text{C}_2}$.  Nevertheless,
the absolute value of this signal is irrelevant since it depends on
many experimental parameters such as the solution concentration, the
probe sensitivity, and the gain of the amplifier. The relevant quantity
is its intensity relative to a reference signal given by the
observation of the initial state $\rho_{\sf init}$. To get a good
signal-to-noise ratio, each experiment (or {\it scan}) was done several
times and the corresponding experimental data were added.

Moreover, to average over small magnetic fluctuations occurring within
the duration of the whole experiment we interlaced scans of the
reference experiment (i.e., the measurement of the reference signal)
with scans of the actual complete pulse sequence. To increase the
spatial homogeneity of the field over the sample we also have inserted
several automated shimming periods consisting of fine tuning of small
additional coils located around the sample. 


\subsection{Results}
\label{results}

\begin{figure}
\begin{center}
\vspace{0.5cm}
\scalebox{0.65}{\includegraphics{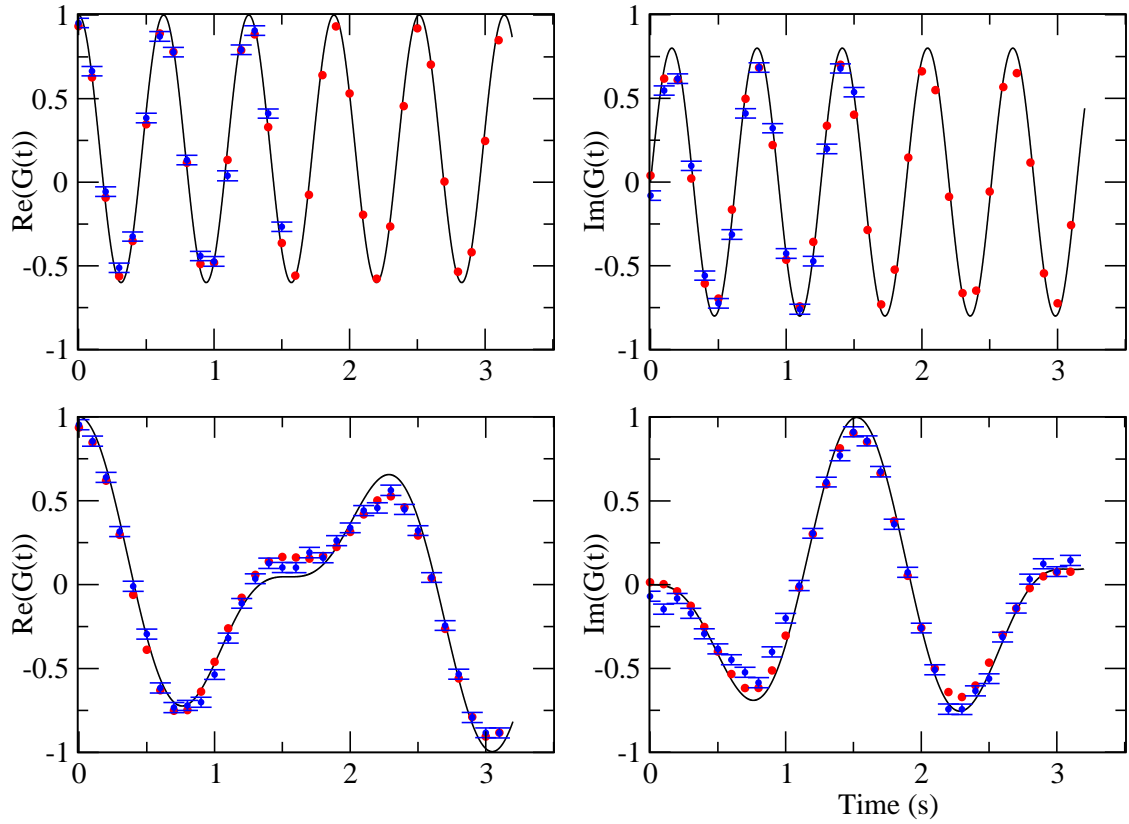}} \end{center}
\caption[Real part of $G(t)$.]{Real and imaginary parts of the
correlation function $G(t)$ of Eq.~\ref{correlation}.  The top panels
show the results when the parameters in Eq.~\ref{Hamilt2} are
$\varepsilon_{k_0} = -2, \epsilon=-8 ,V=4$. The corresponding
parameters $\lambda_1, \lambda_2, \theta$ are in the quantum network,
Fig.~\ref{ei1} are used to measure $G(t)$ and can be determined using
Eqs. \ref{decomp} and \ref{Adecomp}.  The bottom panels show the
results for $\varepsilon_{k_0} = -2, \epsilon=0 , V=4$.  The (black)
solid line is the analytic solution, the red circles are obtained by
the numerical simulation (including the refocusing pulses), and the
blue circles with the error bars are experimental data.}
\label{rescorr1}
\end{figure}

\emph{Correlation function:} In the first experiment we measured the
correlation function $G(t)$ (Eq. \ref{correlation}) for two different
sets of parameters in the Hamiltonian of Eq. \ref{Hamilt2}:
$\varepsilon_{k_0}=-2 \text{, } \epsilon=-8, V=4, $ varying $t$ from
$0.1$ s to $1.5$ s using increments of $\Delta t= 0.1$ s, and
$\varepsilon_{k_0}=-2\text{, } \epsilon=0\text{, } V=4$,   varying $t$
from $0.1$ s to $3.1$ s with $\Delta t= 0.1$ s. The duration of the
optimized pulse sequences from the beginning of the initialization step
to the beginning of the data acquisition, was 97 ms.  In Fig.
\ref{rescorr1} we show the analytical form of $G(t)$ \cite{ortiz1}, as
well as the simulated and experimental data points. The simulated data
points were obtained by a numerical simulation of the Hamiltonian
dynamics of the full seven-qubit register under the optimized pulse
sequence. This simulation is of course inefficient but still
tractable on a conventional desktop computer.

\emph{Hamiltonian spectrum:} 
In the second experiment we measured the function $S(t)$ of Eq.
\ref{fanospectrum} to determine the  eigenvalues of Eq.  \ref{Hamilt2},
for $\varepsilon_{k_0}=-2, \epsilon=-8$, and $V=0.5$. The pulse
sequence applied is the one corresponding to the quantum circuit shown
in Fig.~\ref{fa3} with the corresponding refocusing pulses. Its
duration was about 65 ms. We have repeated this experiment for $128$
different values of the parameter $t$ (Eq. \ref{fanospectrum}), from
$t=0.1$~s to $12.8$~s, using increments of $\Delta t= 0.1$~s.

In Fig.~\ref{resspect} we show the analytical, numerically simulated,
and experimental results for the evaluation of $S(t)$.  As mentioned in
Sec.~\ref{simulation}, a DFT needs to be performed in order to extract
the corresponding eigenvalues.  In Fig.~\ref{fourier} we show the DFT
of the experimental data (see Appendix \ref{app1}), which reveals the
expected peaks at the frequency of the two  eigenvalues of Eq. 
\ref{Hamilt2} in the one-particle sector, for the above parameters.

\emph{Discussion:} At the experimental points, the error bars depend
directly on the signal-to-noise ratio of our experimental data, as it
is obtained after a fit to the experimental measured FID. They can then
be reduced simply by running more scans for each experiment. All 
presented results have been obtained after 8 scans.

Two different classes of errors affect the accuracy of the experimental
results. The first, {\it purely experimental}, type of error is due to
the finite accuracy of the spectrometer, and the intrinsic decoherence
of the physical system we are working with.  The second type of error
is due to the incomplete refocusing induced by the numerical
optimization scheme we used to optimize the pulse sequence. The
numerical simulation of the optimized pulse sequence includes the
errors of the second class but does not take into account the purely
experimental ones. Thus, in our case, the good agreement between
experimental results and simulations suggests that the main
contribution to errors comes from the incomplete refocusing in the
optimization procedure. Increasing the number of refocusing pulses
might have led to more accurate results but would have increased the
overall duration of the pulse sequences. The good agreement between
experiment and simulation is consistent with the fact that the current
duration of the pulse sequences are much smaller than the relevant
relaxation time of the system ($T_2^*$).

\section{Conclusions}
\label{conclusions}

We have successfully simulated a quantum many-fermion system using a
liquid-state NMR based QIP. The algebraic mapping of the operators
describing {\it any} anyonic system onto the Pauli operators describing
our QIP, combined with indirect measurement techniques, allow us to
design efficient algorithms to simulate  arbitrary evolutions of
many-body anyonic systems. 

In this work the system studied was the fermionic Fano-Anderson model,
which can be mapped onto a two-qubit system by use of the standard
Jordan-Wigner transformation. Relevant dynamical correlation functions
of the form $G(t)=\bra{\phi}T^\dagger A_i T B_j \ket{\phi}$  can be
obtained by executing quantum algorithms based on indirect quantum
measurements, i.e., using an additional ancilla qubit. Then, the
algorithm needed to simulate this particular system requires three
qubits.  We were able to design and run pulse sequences to implement
those algorithms on an NMR QIP based on the trans-crotonic acid
molecule (a seven-qubit quantum register). The results obtained agree with
the theoretical ones within efficiently controlled errors. To keep a
constant error level, each pulse sequence has been transformed such
that the time parameters $t_i$ enter as a phase dependence. To shorten
the duration of the pulse sequence and decrease the effect of
decoherence we used only an approximate refocusing scheme. We
numerically optimized those pulse sequences to minimize the error they
introduce in the quantum simulation. These techniques allowed us to get
very accurate results with efficiently controlled errors, since the
overall duration of the pulse sequence was much smaller than the
decoherence time of the system. 

Although the addition of particle-particle (e.g., density-density or
exchange) interactions in the Fano-Anderson Hamiltonian makes it, in
general, non-integrable, the quantum simulation of $G(t)$ remains
efficient, i.e., with polynomial complexity. We can therefore conclude
that this work constitutes an experimental proof of principle for
efficient methods to simulate quantum many-body systems with quantum
computers. 

We thank J. Gubernatis for useful discussions on this subject.
Contributions to this work by NIST, an agency of the US government, are
not subject to copyright laws.

\appendix 

\section{Discrete Fourier Transform and Propagation of Errors}
\label{app1}

\begin{figure}
\begin{center}
\scalebox{0.55}{\includegraphics{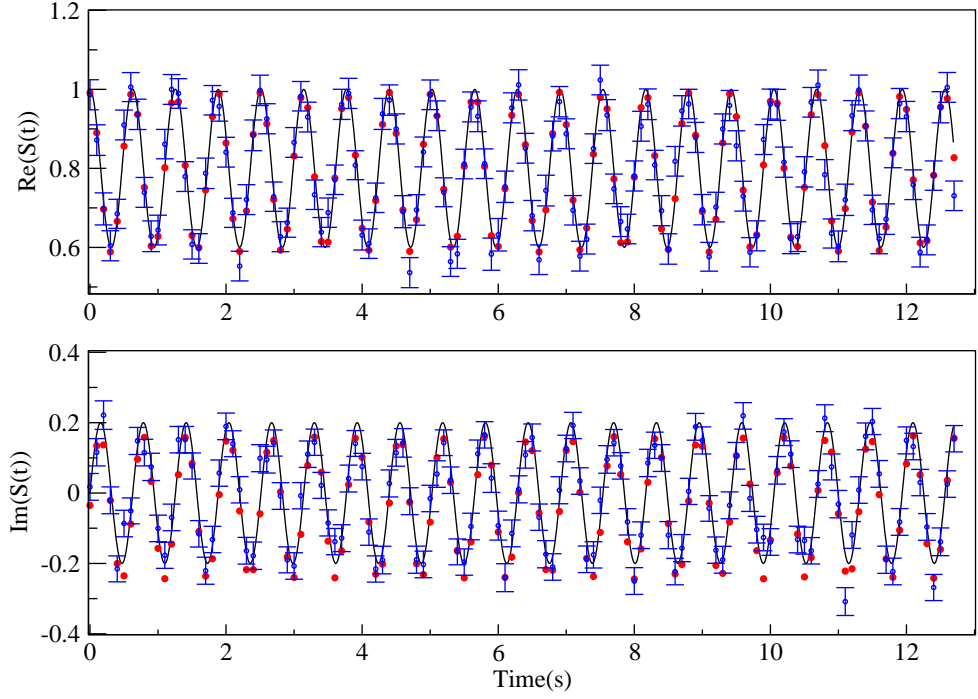}}\end{center}
\caption{Real and imaginary parts of $S(t)$, for 
$\varepsilon_{k_0}=-2, \epsilon=-8$, and $V=0.5$ in Eq.
\ref{Hamilt2}. The (black) solid line corresponds to the analytic
solution. The red circles correspond to the numerical simulation (using
refocusing pulses) and the blue circles with the error bars are
experimental data. $S(t)$ has been measured using the network of Fig.
\ref{fa3} with $\alpha = (\epsilon +\varepsilon_{k_0})/2$. }
\label{resspect}
\end{figure}

Theoretically, the function $S(t)$ of Eq. \ref{fanospectrum} is a
linear combination of two complex functions having different
frequencies: $S(t) = |\gamma_1|^2 e^{-i \lambda_1 t} + |\gamma_2|^2
e^{-i \lambda_2 t}$, where $\lambda_i$ are the eigenvalues of the
one-particle eigenstates, defined as $\ket{1{\sf P}_i}$,  in the
Fano-Anderson model with $n=1$ site and the impurity (see Sec.
\ref{fanoanderson}), and $\lambda_i=|\langle \phi \ket{1{\sf P}_i}|^2$
(Sec. \ref{simulation}), with $\ket{\phi}=\ket{\downarrow_1
\uparrow_2}$ \cite{somma1}. However, the liquid NMR setting used to
measure $S(t)$ experimentally adds a set of errors that cannot be
controlled, and the function $S(t)$ shown in Fig.~\ref{resspect} is no
longer a contribution of two different frequencies only.

\begin{figure}
\begin{center}
\includegraphics[height=7.cm]{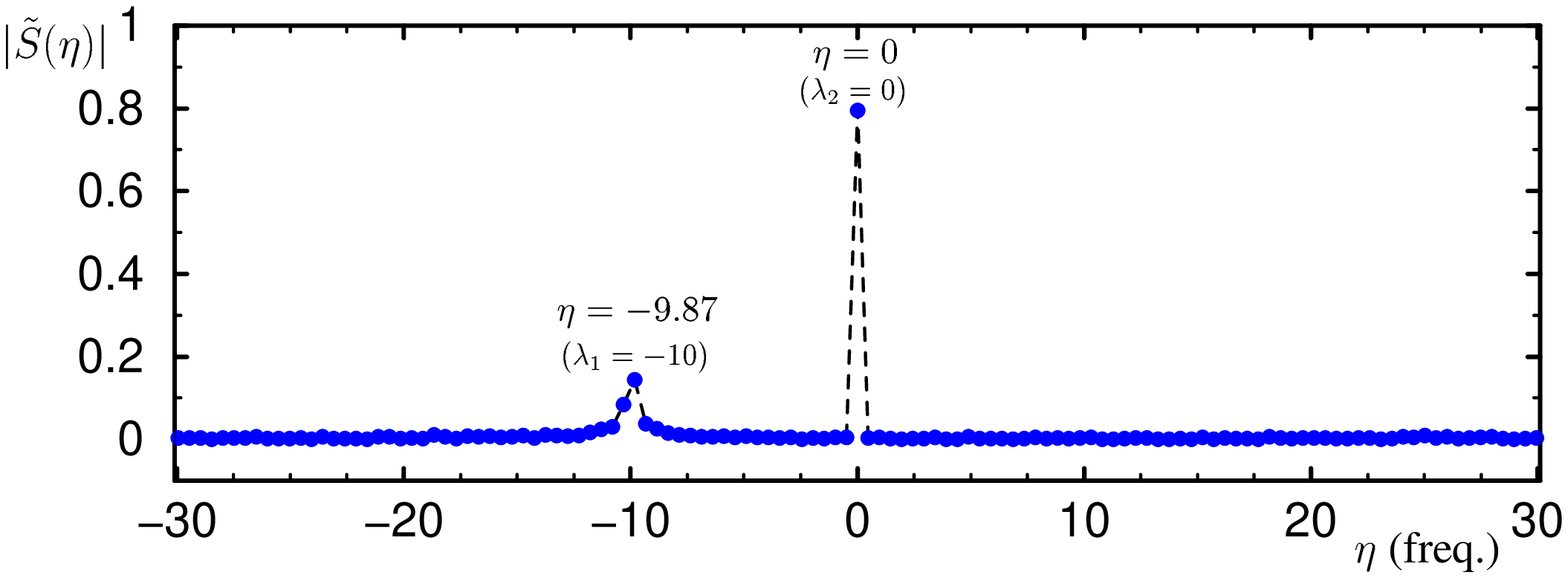}
\end{center}
\caption{Discrete Fourier transform of the real part of the
experimental data of Fig.~\ref{resspect}.  The position of the two
peaks corresponds to the two eigenvalues of the Hamiltonian of Eq.
\ref{Hamilt2} for $\varepsilon_{k_0}=-2, \epsilon=-8$, and $V=0.5$.
Numbers in parentheses denote the exact solution. The size of the dots
representing experimental points is the error bar (see Appendix
\ref{app1}). An upper bound to the error in the frequency domain is
$\approx 0.5$,
which was determined by the resolution of the spectrum.}
\label{fourier}
\end{figure}

As mentioned in Sec. \ref{results}, $S(t)$ was obtained experimentally
for a discrete set of values $t_j = j \Delta t$, with
$j=[1,\cdots,M=128]$ and $\Delta t= 0.1$ s. Its DFT is given by
\begin{equation}
\label{fouriertransform}
\tilde{S}(\eta_l) = \frac{1}{M} \sum\limits_{j=1}^M S(t_j) e^{i \eta_l
t_j},
\end{equation}
where $S(t_j)$ is the experimental value of $S(t)$ at time $t_j$, and
$\eta_l = \frac{2 \pi l}{M \Delta t}$ (with $l=[1,\cdots,M]$) are the
discrete set of frequencies that contribute to $S(t)$ \cite{DFT1}.
Notice that since we are evaluating the spectrum of a physical
(Hermitian) Hamiltonian, the imaginary part of $\tilde{S}(\eta_l)$ is
zero \cite{DFT2}.  In Fig.~\ref{fourier} we show $\tilde{S}(\eta_l)$
obtained from the experimental points $S(t_j)$ of Fig.  \ref{resspect}.
Its error bars (i.e., the size of the line in the figure) were
calculated by considering the experimental error bars of $S(t_j)$ in
the following way: First, we rewrite Eq. \ref{fouriertransform} as
\begin{equation}
\tilde{S}(\eta_l) = \sum\limits_{j=1}^M Q_{lj} ,
\end{equation}
with $Q_{lj} = M^{-1} [{\sf Re}(S(t_j)) \cos( \eta_l t_j) - {\sf
Im}(S(t_j)) \sin( \eta_l t_j)]$ (real).  Then, the approximate standard
deviation ${\sf E}\tilde{S}_l$ of $\tilde{S}(\eta_l)$ depends on
the errors ${\sf E}Q_{lj}$ of $Q_{lj}$ as (considering a normal
distribution \cite{tay1})
\begin{equation}
\label{totalerror1}
[{\sf E}\tilde{S}_l ]^2 \approx  \sum\limits_{j=1}^M [{\sf E}Q_{lj}]^2 .
\end{equation}
On the other hand, ${\sf E}Q_{lj}$ is calculated as \cite{tay1}
\begin{equation}
\label{Qerror}
[{\sf E}Q_{lj}]^2 = \left| \frac{\partial Q_{lj}}{\partial{\sf
Re}(S(t_j))} \right|^2 {\sf E_R}^2 + \left|\frac{\partial
Q_{lj}}{\partial{\sf Im}(S(t_j))} \right|^2 {\sf E_I}^2 ,
\end{equation}
where ${\sf E_R}$ and ${\sf E_I}$ are the standard deviations of the real and
imaginary parts of $S(t_j)$ (see Fig.~\ref{resspect}),
respectively. Because of experimental reasons (Sec. \ref{protocol})
these errors are almost constant, having ${\sf E_R}\sim{\sf
E_I}\sim{\sf E_S}$ independently of $t_j$ (see Fig.  \ref{resspect}),
where ${\sf E_S}$ is taken as the largest standard deviation.  Combining
Eqs. \ref{totalerror1} and \ref{Qerror}, we obtain
\begin{equation}
\label{totalerror}
{\sf E}\tilde{S}_l = \left [ M^{-2} {\sf E_S}^2 \sum\limits_{j=1}^M
[|\cos (\eta_l t_j)|^2 + |\sin (\eta_l t_j)|^2 ] \right] ^{1/2} =
\frac{{\sf E_S}}{\sqrt{M}}.
\end{equation}
In our experiment, $M=128$ and ${\sf E}_S \approx 0.04$, obtaining
${\sf E}\tilde{S}_l \approx 0.0035$, which
determines the (constant) error bars (i.e., the size of the
dots representing data points) shown in Fig.~\ref{fourier}. 

\begin{figure}[hbt]
\label{refoc}
\begin{center}
\includegraphics[height=7.5cm]{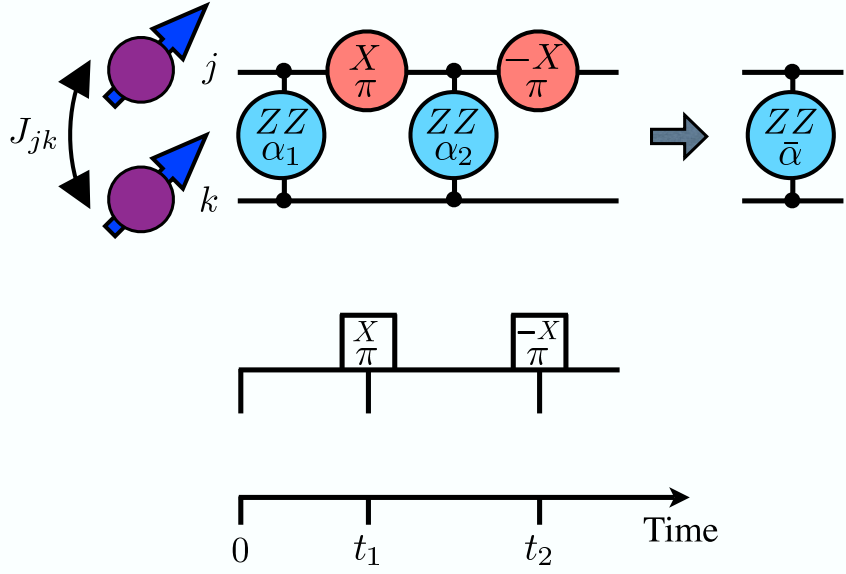}
\end{center}
\caption{Circuit representation for the refocusing scheme to control
$J$-couplings. The Ising-like coupling $J_{jk}$ between spins can be
controlled by performing flips on one of the spins at times $t_1=\Delta
t_1$ and $t_2=t_1+ \Delta t_2$, respectively.  The effective coupling
is $\bar{\alpha} = \alpha_1-\alpha_2 = J_{jk} (\Delta t_1 - \Delta
t_2)$, and vanishes when $\Delta t_1= \Delta t_2$.}
\label{refocalisation}
\end{figure}

The standard deviation ${\sf E}\eta_l$
in frequency domain is due to the resolution of the sampling
time $\Delta t$. This resolution is related to the error coming from the
implementation of the $z$-rotations in the refocusing procedure (Fig.
\ref{refocalisation}). 
A bound for this error is given by the resolution of the spectrum;
that is,
\begin{equation}
{\sf E}\eta_l \le \frac{2 \pi}{M \Delta t}\approx 0.5 \ .
\end{equation}




\newpage

\newpage

\newpage

\newpage

\newpage

\newpage

\newpage

\newpage

\newpage

\newpage

\newpage

\newpage

\newpage

\newpage

\newpage

\newpage


\begin{thebibliography}{10}
\bibitem{note1}
In Feynman's words,  ``If you had discrete quantum systems, what other
discrete quantum systems are exact imitations of it, and is there a
class against which everything can be matched? I believe it's rather
simple to answer that question and to find the class, but I just
haven't done it.''
\bibitem{nota1} 
We actually restrict ourselves to 90 and 180 degrees rotations for
experimental calibration issues.
\bibitem{psudop}
Note that, even though efficient techniques to prepare a pseudo-pure
state exist in theory \cite{refei5}, in practice they are very hard to
implement and one instead uses non-efficient methods that suffer a exponential
decay of observed signal with respect to the number of qubits in the
pseudo-pure state.
\bibitem{had}
The Hadamard gate is the single-qubit rotation $i e^{i \frac{\pi}{2} \sigma_x}
e^{i \frac{\pi}{4} \sigma_y}$.
\bibitem{idealm}
By ideal we mean without experimental error and decoherence effects.
\bibitem{DFT1}
Only a discrete set of frequencies can be obtained from the evaluation
of the DFT over a discrete sample ($S(t_j)$). In this case, a (Nyquist) 
critical frequency is given by $\eta_c = \frac{2 \pi}{\Delta t}$ such that if
$S(t)$ contains frequencies greater than $\eta_c$, they won't be obtained by
applying a DFT (sampling theorem). Notice that two different frequencies
$\eta_l$ and $\eta_{l'}$ are equivalent in a DFT  whenever they differ by an
integer number of $\eta_c$: $\eta_l \equiv \eta_{l'} \Leftrightarrow \eta_l =
\eta_{l'}+ p \eta_c$, with $p$ an integer number.
\bibitem{DFT2}
Due to experimental errors, the imaginary part of $\tilde{S}(\eta_l)$ could be
in principle different from zero. However, we consider only its real part because
it contains all the desired information  (e.g., eigenvalues).
\end{thebibliography}
\end{document}